\documentclass[a4paper,12pt]{article}
\usepackage{latexsym, amssymb, amsmath}
\usepackage{graphicx}
\def \bea{\begin{eqnarray}}
\def \bee{\end{eqnarray}}
\def \be{\begin{equation}}
\def \ee{\end{equation}}
\def \p{\psi}
\def \pnn{\psi_{n+1}}
\def \pn{\psi_n}

\title{Resonance Effects in Topological Discrete sine-Gordon System}
\author{V Kotecha \\ \\ \\ Department of Mathematical Sciences,
University  of Durham, \\ South Road, Durham DH1 3LE, UK}

\begin{document}
\maketitle
\begin{abstract}
We consider kink-antikink collisions in the TDSG system introduced by
Speight and Ward in 1994. We find that the TDSG kink supports extra
internal modes of vibration and  this results in resonance effects of the kind seen for the continuum $\phi^4$ system.
\end{abstract}
\newpage
\section*{ 1.  Introduction}

Nonlinear systems which have soliton solutions have played an increasingly 
important role in physics. One such area in physics is condensed
matter theory where the soliton has been used to model crystal
discolations, charge density waves, magnetic domain walls, etc \cite{cond}. In
many cases it has been  necessary to model interactions of the soliton  
with other objects such as impurities, defects, phonons, antisolitons,
or other solitons. In (1+1)-dimensional field theories, these 
interactions have sometimes been quite spectacular, particularly if the
1-dimensional soliton, or the kink, supports extra modes of
vibration. Such a mode can be excited during an interaction and this
leads to a ``resonance effect'' \cite{camp}. This effect
has been seen in both continuous and discrete systems [1-8]. These 
resonance effects give rise to ``windows'' in the space of impact
velocity, i.e. there are 
certain intervals of velocities in this space for which the kinks form a bound
state (a bion) and others for which the kinks scatter off each other. 

This paper deals with kink-antikink collisions in the topological
discrete sine-Gordon system (TDSG) introduced by Speight and Ward in
1994. This system is particularly interesting since it is a discrete
system but with many of the features of the continuum sine-Gordon. In
particular,  it maintains the Bogomoln'yi bound. The resulting Bogomoln'yi 
equation is a first-order difference equation which can be easily
solved. The solution is analogous to the well-known continuum
sine-Gordon static kink solution. Moreover, as we show in section 4,
the TDSG kink supports an internal shape mode. 

The paper is divided into 5 sections. Section 2 contains a brief
summary of kink dynamics in the TDSG system \cite{sw}, \cite{zak}.  In
section 3 we consider the kink-antikink  
collisions in the TDSG. In section 4 we show that TDSG kink supports
an internal mode of vibration and section 5 contains some concluding
remarks.

\section*{2.  Kink dynamics in TDSG}

The TDSG system is defined by the Lagrangian density

\begin{equation}
\mathcal{L} = \frac{h}{4}\sum_{n\in{Z}}(\dot{\psi_n}^2 - (D_n^2 +
F_n^2)) \label{lag}, 
\end{equation}
where the kinetic energy is $E_K=\frac{h}{4}\sum \dot{\psi}_n^2$ and the
potential energy is $E_P=\frac{h}{4}\sum(D_n^2 + F_n^2)$. 

The quantities $D_n$ and $F_n$ are given by  

\bea
D_n=\frac{2}{h}\sin\frac{1}{2}(\psi_{n+1} - \psi_n) \quad
\textnormal{and} \quad F_n=\sin \frac{1}{2}(\psi_{n+1} + \psi_n), 
\bee 
giving 

\be
E_P=\frac{h}{4}\sum\frac{4}{h^2}\sin^2\frac{1}{2}(\psi_{n+1}-\psi_n) +
\sin^2\frac{1}{2}(\psi_{n+1}+\psi_n). \label{pot}
\end{equation}
$\psi_n=\psi (x)$ at the $n^{th}$ lattice site and $\psi_{n+1} =
\psi(x + h)$, where $h$ is the lattice spacing.  The first term in
equation (\ref{pot}) represents an attractive force analogous to the Hooke
force. The second term is the substrate potential, found by taking an
average of the potential of the two nearest neighbours. 

In the continuum limit, as $h \rightarrow 0$, $D_n\rightarrow \psi_x$,
and $F_n\rightarrow \sin\psi$, and one recovers the standard
expression of the continuum potential 

\be
E^{cont}_P(\psi)=\frac{1}{4}\int_{-\infty}^{\infty}(\psi_x^2 + \sin^2\psi)\, dx.
\end{equation}
The key feature of this model is the choice of the derivative on the
lattice. This choice is motivated by the Bogomol'nyi argument of the
continuum system. Rather than choose the standard forward difference,
Speight and Ward choose,

\be
- \frac{\cos \pnn - \cos\pn}{h} = D_nF_n,
\end{equation}
with the kink boundary conditions
\begin{displaymath}
\lim_{n\rightarrow - \infty} \pn = 0 \quad \textnormal{and} \quad
\lim_{n\rightarrow \infty} \pn = \pi. 
\end{displaymath}
This leads to the following discrete Bogomol'nyi argument

\begin{align}
0 &\leq \frac{h}{4}\sum_{n\in  Z}(D_n - F_n)^2 \\
&= E_P + \frac{1}{2} \sum_{n\in Z}(\cos\pnn - \cos\pn) \\
&= E_P - 1, 
\end{align}
so $E_P \geq 1$, with equality if $D_n=F_n$,
i.e. if

\begin{align}
\qquad \qquad \frac{2}{h}\sin\frac{1}{2}(\psi_{n+1} - \psi_n) \quad &= \sin \frac{1}{2}(\psi_{n+1} + \psi_n) \\
\textnormal{or,} \quad \quad \pn  &= 2 \tan^{-1} \exp{a(nh-b)} \label{kink}.
\end{align}
The constant $a=h^{-1}\ln[\frac{2+h}{2-h}]$ is the kink's slope and $b\in R$
is the position. Equation (\ref{kink}) is the static kink solution. So the
moduli space of static solutions for the TDSG is isomorphic to
$\mathbb{R}$ rather than $\mathbb{Z}$, which is what one usually expects
for a discrete system. Moreover, $E_P$ is independent of $b$
suggesting that the model has no Peierls-Nabarro barrier. This means
that the kink can move along the lattice arbitrarily slowly without
getting pinned. 

The Euler-Lagrange equation gives the following equation of motion,

\be
\ddot {\psi_n} = \frac{4-h^2}{4h^2} \cos \pn(\sin \pnn + \sin
\p_{n-1}) - \frac{4+h^2}{4h^2}\sin \pn (\cos \pnn + \cos \p_{n-1}). \label{eom}
\end{equation}
Equation (\ref{eom}) can then be used to study the behaviour of a moving
kink. Speight and Ward have solved (\ref{eom}) using a 4th-order Runge-Kutta
scheme. Their initial condition was a static kink Galilean boosted to
a velocity $v$. Their results show that for small initial velocities
the kink wobbles with a period $h/v$ as it moves along the
lattice. This is a purely dynamical effect since there is no
Peierls-Nabarro barrier. 

For fast moving kinks it is found that the kink radiates as it moves
along the lattice. This is observed through a gradual decrease in the
kink velocity. But it is found that there is certain velocity
threshold below which the radiation from the kink is significantly
reduced.  

The existence of such a velocity threshold can be understood by
considering the linearised equation of motion \cite{sw}. 

\be
\ddot{\psi_n} = \frac{4-h^2}{4h^2}(\pnn + \p_{n-1}) -
\frac{4+h^2}{2h^2}\psi_n. \label{lin}
\end{equation}
Equation (\ref{lin}) can be used to derive a dispersion relation for small
amplitude travelling waves, given by 

\be
\omega^2 = \frac{4 +h^2}{2h^2} - \frac{4-h^2}{2h^2}\cos kh,
\end{equation}
where $k$ is the wavenumber. Since $0<h<2$, we have
$1<\omega<2/h$. The case $\omega = 1$ corresponds to the threshold
velocity. The frequency at which the kink hits the lattice sites is
$v/h$ per unit time. So provided $v\geq h/2\pi$, the kink will radiate
(at $\omega = 2\pi v/h$). But if $v< h/ 2\pi$ then $\omega<1$
and the kink does not radiate. Hence there is a "preferred" velocity
for the kink. Since this velocity depends only on $h$, it can be
regarded as a feature of the lattice. 

The choice of the functions $D_n$ and $F_n$ is not unique. For
instance, $D_n$ could be multiplied by a function $f(x,h)$ and $F_n$ by
$f(x,h)^{-1}$ and the product is independent of $f$ so one would still
have a topologically stable kink solution. The only condition on
$f(x,h)$ is that $\lim_{h\rightarrow 0} f=1$. Similarly there is
freedom in the choice of $\ddot \psi$. Zakrzewski \cite{zak} has
investigated the effects on the kink motion when this choice is
exploited. 

The $\ddot \psi$ in (\ref{eom}) is replaced by 

\be
\ddot \psi \rightarrow \frac{\ddot \psi}{1+\alpha  \dot \psi^2}
\end{equation}
where $\alpha=\alpha(h)$ and $\alpha(0)=1$. The equation of motion (\ref{eom}) is now changed to

\be
\frac{\ddot \psi_n}{(1 + \alpha \dot{\psi_n}^2)} =
\frac{f^2}{h^2}(\sin (\pnn - \psi_n) - \sin(\psi_n - \p_{n-1})) -
\frac{1}{4f^2}(\sin(\pnn + \psi_n ) + \sin(\psi_n + \p_{n-1})). \label{modeom}
\end{equation}
The choices of $f$ and $\alpha$ are motivated by the fact that equation (\ref{modeom}) still needs to admit the Galilean boosted sine-Gordon kink solution. One choice is the following

\begin{equation}
f^2 = \frac{\frac{v_1^2}{1-v_1^2} +
\sqrt{\left(\frac{v_1^2}{1-v_1^2}\right)^2 + \frac{4\sinh^2
2\beta}{h^2}}}{\frac{8 \sinh^2 \beta}{h^2}} 
\end{equation}

where $\beta = h/2\sqrt{1-v_1^2}$, and 

\begin{equation}
\alpha(h) = \frac{1-v_1^2}{v_1^2}\sinh^2(2\beta).
\end{equation}

Numerical simulations of (\ref{modeom}) using a 4th-order Runge-Kutta method has
shown that the kink propagates without emitting any radiation if the
initial velocity of the kink is chosen to be the velocity $v_1$ used
in the expression of $f^2$. For initial velocities $v\leq v_1$, there
is virtually no emission of radiation so the kink velocity hardly
decreases. For $v > v_1$, the kink initially propagates at velocity
greater than $v_1$ but then slows down to a velocity just less than $v_1$
after which it's virtually constant with little emission of
radiation. The initial increase in velocity is due to the readjustment
of the kink configuration to match the lattice field configuration. 

\section*{3.  Kink-antikink Collisions in the TDSG System}

\subsection*{3.1  Preliminaries}

In this section we present results of kink-antikink interactions in
the Speight-Ward model. The results are obtained by solving (\ref{eom})
with the initial condition corresponding to Galilean boosted static
kink and antikink solutions at each end of the lattice grid, i.e. 

\begin{align}
\psi(x,0) &= 2\tan^{-1}\exp a(x + b) - 2\tan^{-1}\exp a(x - b), \\
\dot{\psi}(x,0) &= -av(\textnormal{sech} \, a(x+b) + \textnormal{sech}
\, a(x-b)). \label{initial} 
\end{align}
The boundary condition is taken to be $\psi(x_{min}) = \psi(x_{max}) =
0$, i.e. the ends of the kink and anti-kink are held fixed at all
times.  

The kink and antikink move towards each other with relative velocity
$2v$. Equation (\ref{eom}) is solved using a 4th-order Runge-Kutta
algorithm. The programme is run with various lattice spacings $h$.
The time step is chosen to be 0.05. This conserves the energy to
within 0.05\% in a simulation which runs for 1000 units. We use the
same time step for all values of $h$.  

Our choice of the boundary condition means that any radiation emitted
by the kink will reflect off the edges and eventually interact with the
kinks. This is easily avoided by making the grid sufficiently
large.

As mentioned already, the amount of radiation emitted by the kink
depends on the kink velocity and also on the lattice spacing. For large
velocities and coarser lattices the amount of radiation is
significantly greater. In these cases 
we have damped the first few lattice sites at each end of the
grid. That is, the value of $\dot{\psi}$ at these
sites is decreased by a constant amount (10\%) for the entire duration of the
simulation. The total energy of the system is of course no longer
conserved. The energy loss depends on the initial velocity of the kink
and on the lattice spacing, but it is usually within 10\% of the
initial energy, for a simulation running for 1000 units. 

\subsection*{3.2 Simulation Results} 

The simulations showed that the outcome of an interaction depends on
the impact velocity of the kinks. The velocity of the kink (or the antikink) is defined to be the rate of change of $X(t)$, where $X(t)$ is the ``average'' position of the kink (or the antikink). $X(t)$ for the kink is given by

\begin{equation}
X(t)
= \frac{ \displaystyle \sum_{i = -N}^0 x_ie_i}{ \displaystyle \sum_{i = -N}^0 e_i}.
\end{equation}
Here $e_i$ is the total energy of the field and the radiation at the $i^{th}$ lattice site, where $x_i = x_{min} + ih$. The quantity $\sum_{-N}^0 e_i$ is the total energy of the field and the radiation for the single kink. We found that for cases when the phonon radiation from the kinks is not significant, $X(t)$ makes a good approximation to the kink position. We define the velocity of the kink to be $\dot{X}(t)$, which is computed using the forward difference, $\dot{X}(t) = \frac{1}{\Delta t}(X(t+\Delta t) - X(t))$. 

Since the kink wobbles as it moves along the lattice, $\dot{X}(t)$ is
an oscillating function. The period of oscillation is the wobble
period, $h/v$, where $v$ is the velocity used in the initial condition
(\ref{initial}). This wobble can be understood by considering the geodesic motion
on the kink moduli space. This moduli space has a periodic metric \cite{sw}. 

Figure 1 shows various plots of kink velocity against time for kinks
with different initial positions. The kinks are on a lattice with
$h=1.4$ and the initial velocity of the kink is $v=0.197$. In all
cases the average velocity of the kink is significantly
decreased. This is due to phonon radiation from the kinks. The plots
also show that the kink velocity depends on the initial position of
the kink, modulo $nh$, where $n \in \mathbb{Z}$. 

The outcome of an interaction has sensitive dependence on the velocity
of the kink. In particular, there are three different outcomes:

\emph{(a)} $\,$ If the kink impact velocity is above a certain
critical velocity, 
$v_c$, then the kinks simply pass through each other, i.e. there is a
smallest number $v_c$ such that the outcome of the interaction for all
$v \ge v_c$ is always a passing through behaviour. The value of $v_c$
depends on the lattice spacing and also on $b$. Figure 2 shows the
dependence of $v_c$ against $h$. The dotted curve represents the
quadratic $0.073h^2+0.036h$. This curve is an empirical fit to the
data points and is not derived from the theory. 

The dependence of $v_c$ on the position of the kinks is due to
dynamical dressing and the kink wobble. For small initial velocities
(figure 1), the kink velocity oscillates with a period $(h/v)$. The
amplitude of oscillation is of the order of $10^{-2}$. For large
velocities, the phonon radiation make the oscillations erratic and
also significantly decrease the overall velocity of the kink. The
initial velocity also changes due to the readjustment of the kink
field to suit the lattice distribution (dynamical dressing). The
initial propagating velocity of the kink is not always $v$, the
velocity used in expression (\ref{initial}). In some cases it is
larger than $v$ 
and in others it is less than $v$. This change depends on the
magnitude of $v$ and on the lattice spacing. The effect can be seen in
figure 1. 
\begin{figure}
\begin{center}
\includegraphics[angle=270, width=0.7\textwidth]{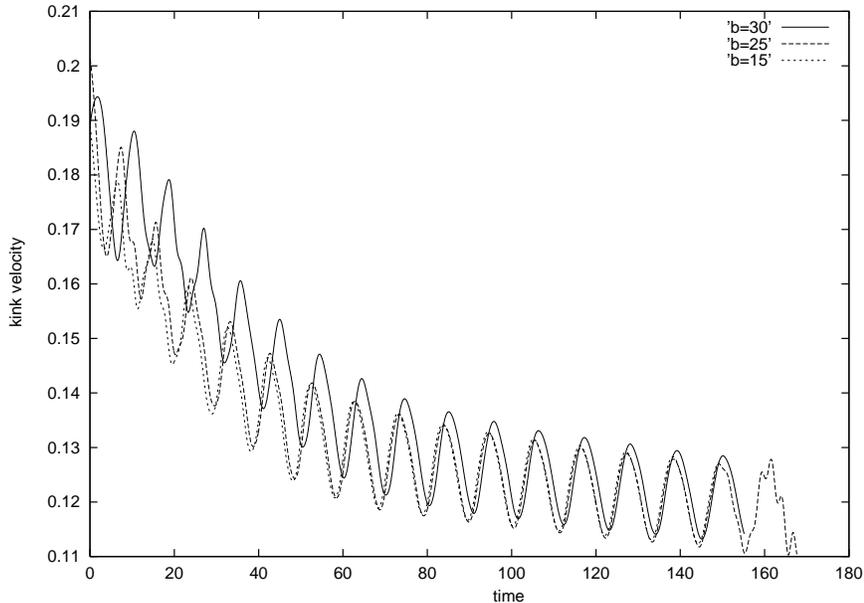}
\caption{A plot of the kink velocity against time. The three curves
represent velocities of kinks with an initial velocity of 0.197 but
starting at different points on the lattice. The solid curve
represents a kink with $b=30$, the dashed curve a kink with $b=25$ and
the dotted curve is for $b=15$.} 
\label{fig: velo_pos} 
\end{center}
\end{figure} 

\begin{figure}[h]
\begin{center}
\includegraphics[angle=270, width=0.7\textwidth]{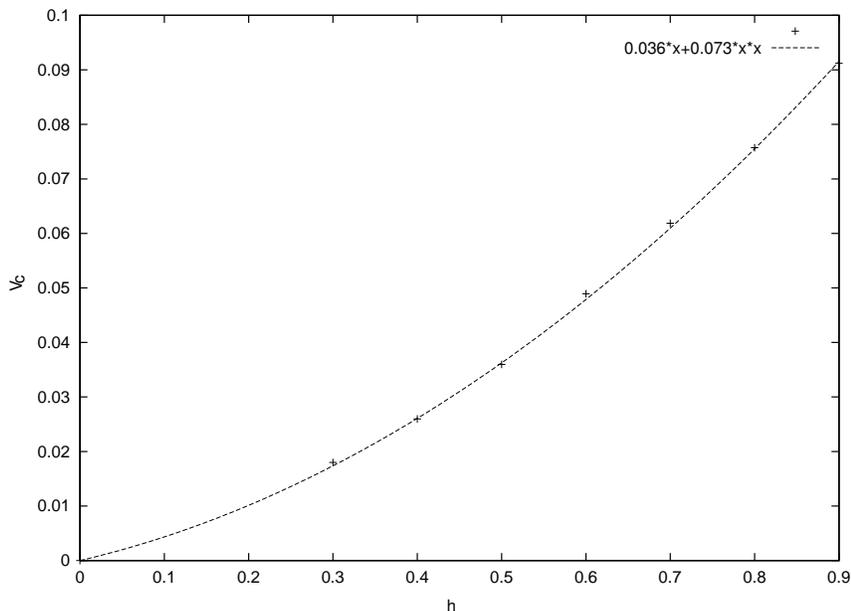}
\caption{A plot of critical velocity against lattice spacing. The
crosses represent data points found numerically. The curve represents
a theoretical fit of $v_c = 0.073h^2 + 0.036h$.} 
\label{fig: v_c} 
\end{center}
\end{figure}

For initial kink velocities which give impact velocities below $v_c$,
we found that the outcome is either $\emph{(b)}$ a bounce or
\emph{(c)} a trapped
breather-like state (a bion). The bounce is essentially a
kink-antikink scattering, where the kinks after the first collision
pass through each other but fail to escape to infinity. Instead, they
travel a small distance and then turn around for another
collision. After the second collision the kinks either escape to
infinity or turn around again for more collisions. For a 2-bounce
event (meaning the kinks pass each other twice) the ultimate outcome of this scattering would be a reflection, and for a 3-bounce a transmission.  
 
The velocities for the bounces occur in small intervals. For
velocities outside of these intervals the outcome is always a bound
state. This is similar to the results for the $\phi^4$ model \cite{ann},
except in this case the outcome does not depend fractally
on the impact velocity. We have only seen 2, 3 and
4-bounce events. But the velocities at which the bounces occur are not easy to find and the intervals are small ($10^{-7}$ for a 4-bounce),
so it may be that higher bounce events occur but we have not seen
them. 

Figure 3 shows an example of the 2-bounce event. In this figure we
have plotted the field $\psi(0,t)$ against $t$. $\psi(0,t)$ is the
field at the centre of mass of the kink-antikink system.The kinks are on a lattice
of unit spacing with an initial  separation of 10 units. $v_c$ for
this combination is found to be 0.100. The kinks bounce for all
initial velocities in the range $0.0969\leq v \leq 0.0972$ (2-bounce
window).    
 
Figures 4 and 5 show examples of three and four bounce events. In
figure 4 the kinks transmit whereas in 5 they reflect off each
other. As can be seen from the figure for the 2-bounce case, the kink
and antikink collide, pass through each other and travel a small
distance before stopping, turning around and travelling in the other
direction. The kinks then collide again, pass through each other and
this time escape to infinity.  
\begin{figure}
\begin{center}
\includegraphics[angle=270, width=0.7\textwidth]{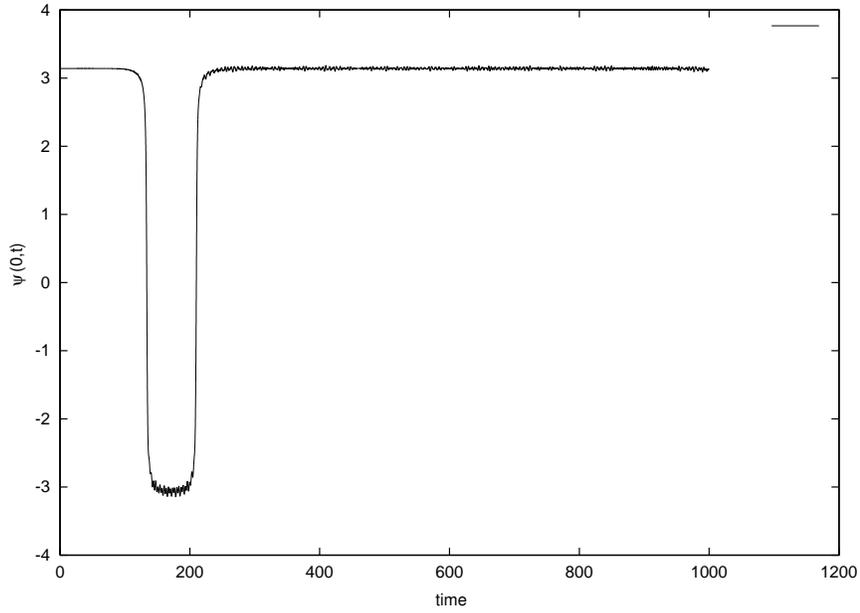}
\caption{A figure representing the 2-bounce event. The kinks collide and essentially reflect off each other.}
\label{fig: 2-bounce}
\end{center}
\end{figure} 
\begin{figure}
\begin{center}
\includegraphics[angle=270, width=0.7\textwidth]{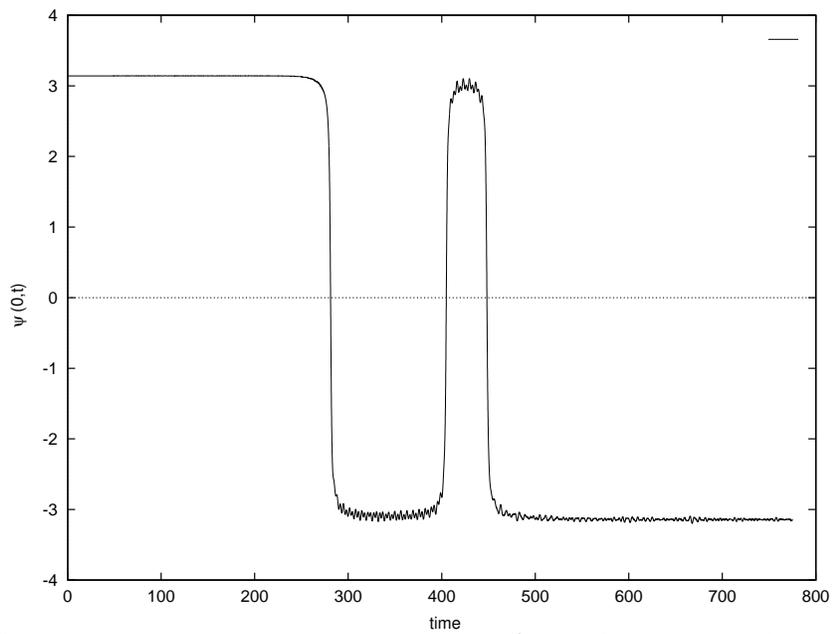}
\caption{A 3-bounce event where the scattering of the kinks results in a transmission.}
\label{fig: 3-bounce}
\end{center}
\end{figure}
\begin{figure}
\begin{center}
\includegraphics[angle=270, width=0.7\textwidth]{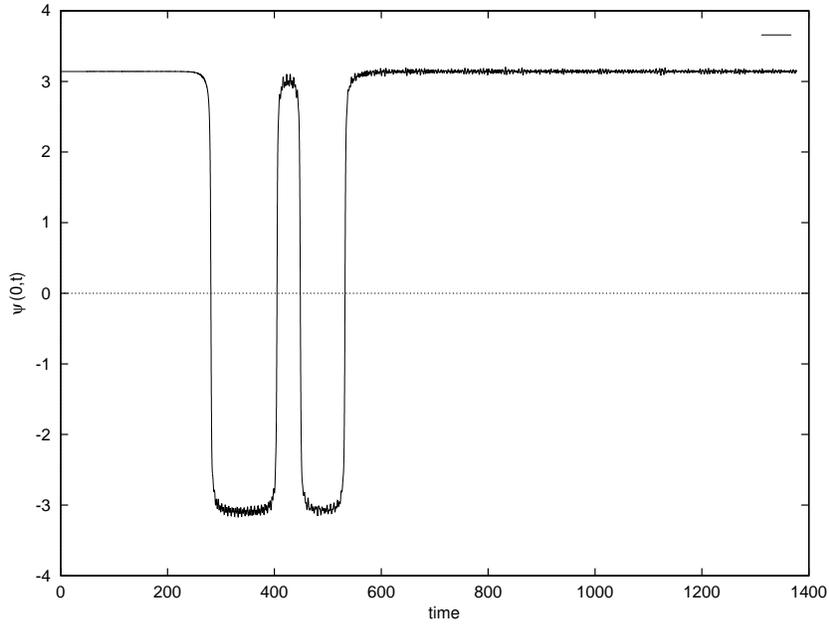}
\caption{A plot showing a 4-bounce event.}
\label{fig: 4-bounce}
\end{center}
\end{figure}
\begin{figure}
\begin{center}
\includegraphics[angle=270, width=0.7\textwidth]{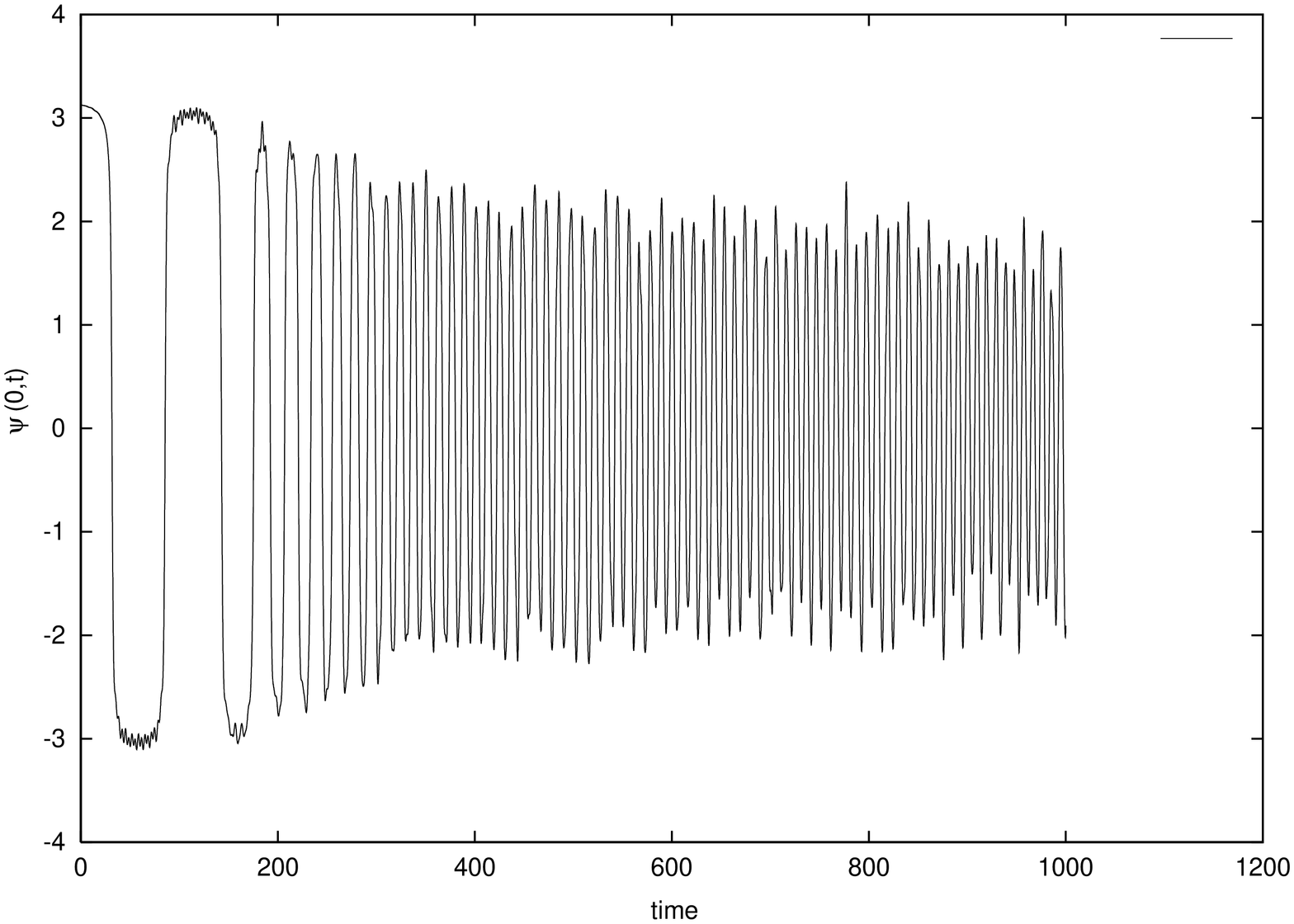}
\caption{A plot of $\psi(0,t)$ against time, representing the
oscillations at the centre of mass of the bion.}
\label{fig:bion}
\end{center}
\end{figure}
Figure 6 is a graph of $\psi(0,t)$ against $t$ representing the
oscillations at the centre of mass of a bion. This is essentially an
$n$-bounce event for $n$ very large. The erratic oscillations suggest
that the system might be chaotic. The period of oscillations become
constant after about 10 cycles and the bion 
then oscillates for a long time (i.e for hundreds of cycles). We have
calculated the maximum Lyapunov exponent of the time series
$\psi(0,t)$, using the algorithm of Wolf \emph{et al}
\cite{wolf}. Its value is approximately 0.18, suggesting that the
oscillations are mildly chaotic. In general, a positive maximum
Lyapunov exponent is taken as a formal definition of chaos.

The results of the simulation can be understood using the resonance
energy exchange mechanism. This requires the TDSG kink to support
shape modes. We show next that the TDSG kink does support such a
mode. The frequency of this mode is calculated in two ways, by using a
collective coordinate approximation and also numerically, by Fourier
analysing the deformations of a perturbed static kink. Note that we
could also have used a matrix-based approach similar to the one used
by Speight in \cite{casimir}, where it was used to calculate the
Casimir energy of the kink. 

\section*{4.  Kink internal shape mode}

\subsubsection*{4.1  Collective coordinate approximation}

We can obtain a collective coordinate approximation of the TDSG system
by treating the scale factor $a$ in the kink as a dynamical variable. This
reduces the number of degrees of freedom from infinity to one. The
Lagrangian of the reduced system is given by 

\begin{equation}
\mathcal{L} = f(a,b)\dot{a}^2 - g(a,b),
\end{equation}
where $f(a,b)$ and $g(a,b)$ are functions given by

\begin{align}
f(a,b) &= \frac{h}{4} \sum_x (x-b)^2 \textnormal{sech}^2 a(x-b), \\
g(a,b) &= \frac{1}{h} \tanh(\frac{1}{2}ah) + \frac{h}{4}\coth(\frac{1}{2}ah).
\end{align}
The corresponding equation of motion is 
\begin{equation}
2f\ddot{a} + f'\dot{a}^2 - g' = 0.
\end{equation}
 where the prime denotes differentiation with respect to $a$. $g(a,b)$ has a minimum (stable equilibrium) at
$a=s:=\frac{2}{h}\tanh^{-1}\frac{h}{2}$. Small amplitude oscillations $a(t) = s+\epsilon(t)$ satisfy the linearised equation, 
\begin{equation}
2f(s,b)\ddot{\epsilon} = g''(s,b)\epsilon.
\end{equation}
Hence the waves oscillate with frequency
\begin{equation}
\nu = \frac{1}{2\pi}\sqrt \frac{g''(s)}{2f(s,b)}.
\end{equation}
 $g''(s)=(1-\frac{1}{4}h^2)^2$, so the frequency of the shape mode
depends on the lattice spacing. For $h=1$, the shape mode frequency is
0.152. Table 1 shows the dependence of frequency on the lattice spacing $h$.
\\

\begin{center}
\begin{tabular}{|r|l|} 
\hline
h & frequency \\
\hline \hline
0.1 & 0.175 \\
0.2 & 0.175 \\
0.5 & 0.170 \\
0.8 & 0.161 \\
0.9 & 0.157 \\
1.0 & 0.152 \\
1.2 & 0.143 \\
\hline
\end{tabular}
\end{center}

As a check on the accuracy of the collective coordinate approach, we
have used it to calculate the frequency of the shape mode of the
$\phi^4$ kink. The value found is $\sqrt(1.55)$, which is fairly close
to the actual value $\sqrt(1.5)$.  

\subsubsection*{4.2  Numerical method}

Numerically, the shape mode frequency is obtained by Fourier analysing the deformations of the
perturbed static kink. We set the initial condition for equation (\ref{eom})
to be   

\begin{align}
\psi(x,0) &= \psi_K(x,0) + \epsilon \\ \nonumber
\dot{\psi}(x,0) &= 0,
\end{align}
where $\psi_K(x)$ is the static kink solution and $\epsilon$ is the
pertubation. The value of $\epsilon$ is chosen to be -0.03. The field 
$\psi(x,t)$ is sampled at various values of $x$. From this we
obtain the variations in the field, $\delta \psi(x,t) = \psi(x,t) -
\psi(x,0)$, for each of the sampled values of $x$. $\delta \psi(x,t)$
were then Fourier analysed using the {\footnotesize MATLAB} FFT
routine.   

In figure 7 we have plotted the power spectrum for the data sampled
at $x=0$ for a kink on an $h=1$ lattice. The power spectrum of the
data for other values of $x$ give identical power spectrums.
\begin{figure}[h]
\begin{center}
\includegraphics[angle=0, width=0.8\textwidth]{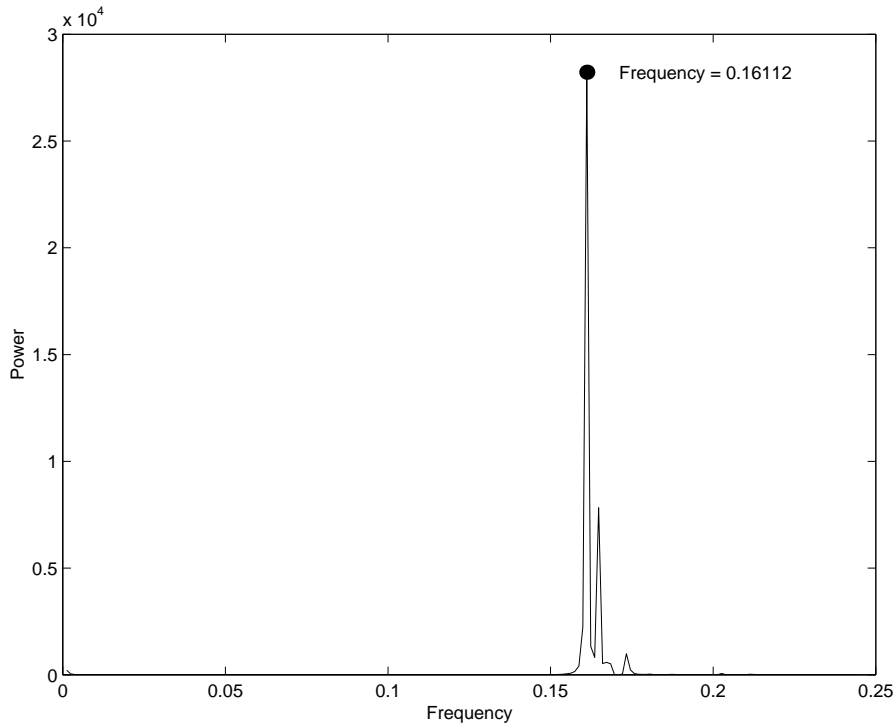}
\caption{Power spectrum of the fluctuations around a perturbed static kink}
\label{fig:power}
\end{center}
\end{figure}   

Figure 7 was constructed using $N=2^{14}= 16384$ data
points. Aliasing of the power does not occur since the amplitude at
the Nyquist frequency is essentially zero. There is a peak at
$\nu=0$ corresponding to the Goldstone mode, and a peak at $\nu =
0.161$ corresponding to the shape mode.   

This procedure is repeated for other values of $h$. In contrast to
the results suggested by the collective coordinate approach, we find
that the frequency of the shape mode is more or less independent of
$h$. For $0.5\le h \le 1.2$, we found $0.158<\nu<0.161$. The value of
the frequency (for $h=1$) however is within the range of frequencies found
using the collective coordinate method. 

In the continuum limit the TDSG shape mode frequency is precisely the
quasimode frequency of the continuum sG system found by Rice,
$\omega_R = 0.175$ \cite{rice}, although it is unlikely that such a
quasimode exists \cite{qsm}  

It is interesting to compare the situation with other discrete
sine-Gordon systems. The well-known Frenkel-Kontorrova model does not
support a kink with an internal shape mode for all values of the
discreteness parameters. The sine-Lattice model (s-L) however does
support one, as long as the system is only weakly discrete
\cite{zhang}. The shape mode for the s-L system lies just above the
lower phonon band suggesting that it is a genuine kink shape mode
rather than just a phonon resonance. Moreover, in the continuum limit
this shape mode converges to the sine-Gordon quasimode frequency. The
existence of a shape mode in a discrete sine-Gordon system depends on
the anharmonicity of the potential term.

\section*{5. Conclusion}

We have found that the kink-antikink interactions in the TDSG system
exhibit resonance phenomena similar to those in the continuum $\phi^4$
system. These are found to be due to the excitation of an intermal mode of the
TDSG kink. A collective coordinate analysis shows that the frequency
of this mode depends on the lattice spacing, and in the continuum
limit $(h \longrightarrow 0)$ the frequency seems to approach the
quasimode frequency predicted by Rice. The recent analysis done in \cite{qsm}
however says that there cannot be a quasimode for the continuum sine-Gordon
kink, but they do predict the existence of a quasimode for a lattice
sine-Gordon kink. This is not the quasimode we have found, so this is
something genuinely new. Further numerical simulations are required to
determine what is happening in the limit $h \longrightarrow 0$. 

Also, as we have mentioned already, there is a modified TDSG system
\cite{zak} which admits an exact travelling-kink solution (with
a fixed velocity). It would be interesting to perform kink-antikink
interactions for this system. Since the expressions we have used to
perform the interactions are only an approximation to the equation of
motion, it would be interesting to see the difference.

We have also done kink-antikink collision simulations for the
TD$\phi^4$ system \cite{speight:phi4}. The results are found to be
analoguos to the conventional discrete $\phi^4$ system \cite{ann},
though in the ``windows'' structure in the space of impact velocity is
not a fractal. 

\emph{Acknowledgements} I would like to thank Richard Ward for helpful
discussions and EPSRC for a research studentship.


\begin{thebibliography}{99}
\bibitem{cond} A.R.Bishop and T.Scheider (1983) \emph{Solitons and
Condensed Matter Physics} (Springer-Verlag).
\bibitem{camp} D.K.Campbell, J.F.Schonfeld and C.A.Wingate
(1983). Resonance structure in kink-antikink interactions in $\phi^4$
theory, \emph{Physica} \textbf{9 D} 1-32.
\bibitem{kud} T.Belova and A.E.Kudryavtsev (1988). Quasiperiodical
orbits in the 
scalar classical $\lambda \phi^4$ field theory, \emph{Physica
D} \textbf{32} 18. 
\bibitem{ann} P.Anninos, S.Oliveira, and R.A.Matzner (1991). Fractal
structure in the scalar $\lambda(\phi^2-1)^2$ model, \emph{Phys Rev}
\textbf{D 44} 1147.
\bibitem{ablowitz} M.J.Ablowitz, M.D.Kruskal and J.F.Ladik (1979). Solitary
wave collisions, \emph{Siam J Appl Maths} \textbf{36} 428-437. 
\bibitem{dsg} V.A.Gani and A.E.Kudryavtsev (1998). Kink-antikink
interactions in 
the double sine-Gordon equation and the problem of resonance
frequencies, cond-mat/9809015. 
\bibitem{zhang} Fei Zhang (1997). Kink shape modes and resonant dynamics in
sine-lattices, \emph{Physica D} \textbf{110} 51-61. 
\bibitem{kur} A.E.Kudryavtsev (1975). Solitonlike solutions for a
Higgs scalar field., \emph{JETP Lett} \textbf{22} 82-83.
\bibitem{c} D.K.Campbell, M.Peyrard and P.Sodano (1986). Kink-antikink
interactions in the double sine-Gordon equation, \emph{Physica}
\textbf{19 D} 165-205. 
\bibitem{sw} J.M.Speight and R.S.Ward (1994). Kink dynamics in a novel
discrete sine-Gordon system, \emph{Nonlinearity} \textbf{7}
475-484.
\bibitem{para} M.Peyrard and D.K.Campbell (1983). Kink-antikink interactions
in the parametrically modified sine-Gordon system, \emph{Physica} \textbf{9 D}
33-51.
\bibitem{rice} M.J.Rice (1983). Physical dynamics of solitons,
\emph{Phys Rev B} \textbf{28} 3587-3589.
\bibitem{bw} R.Boesch and C.R.Willis (1990). Existence of an internal
quasimode for a sine-Gordon soliton, \emph{Phys Rev B} \textbf{42} 2290.
\bibitem{zak} W.J.Zakrzewski (1994). A modified discrete sine-Gordon model,
\emph{Nonlinearity} \textbf{8} 517-540.
\bibitem{wolf} A.Wolf, J.B.Swift, H.L.Swinney and J.A.Vastano (1985).
Determining Lyapunov exponents from a time series, \emph{Physica D}
\textbf{16} 285-317. 
\bibitem{casimir} J.M.Speight (1994). Kink Casimir energy in a lattice
sine-Gordon model, \emph{Phys Rev} \textbf{D 49} 6914-6919.
\bibitem{qsm} Niurka R. Quintero, Angel Sanchez and Franz G. Mertons
(2000). Existence of internal modes of sine-Gordon kinks, \emph{Phys
Rev} \textbf{E 62} R60-R63.
\bibitem{speight:phi4} J.M.Speight (1997). A discrete $\phi^4$ system
without a Peierls-Nabarro barrier, \emph{Nonlinearity} \textbf{10}
1615-1625. 
\end{thebibliography}
\end{document}